\documentstyle[12pt]{article}
\newcommand{\beq}{\begin{equation}}
\newcommand{\eeq}{\end{equation}}
\newcommand{\id}
 {i\kern.06em\hbox{\raise.25ex\hbox{$/$}\kern-.60em$\partial$}}
\newcommand{\as}{/\kern-.52em a}
\newcommand{\bs}{/\kern-.52em b}
\newcommand{\D}{{\cal{D}}}
\newcommand{\dv}{\!d^3\!x\,}
\newcommand{\Z}{{\cal Z}}
\renewcommand{\d}{\partial}
\newcommand{\Fd}{{^*}{\!F}}
\newcommand{\J}{{\cal J}}
\newcommand{\tr}{\mathop{\rm tr}\nolimits}
\begin{document}
\title
{Bosonization of Three Dimensional Non-Abelian Fermion Field Theories}
\author{Ninoslav Brali\'c\\
{\normalsize\it
Facultad de F\'\i sica, Pontificia Universidad Cat\'olica de Chile,}\\
{\normalsize\it
Casilla 306, Santiago 22, Chile.}\\
\rule{0cm}{1.cm}
Eduardo Fradkin\\
{\normalsize\it
Department of Physics, University of Illinois at Urbana-Champaign,}\\
{\normalsize\it
1110 W.~Green St.~, Urbana Illinois 61801-3080, USA.}\\
\rule{0cm}{1.cm}
Virginia Manias\thanks{CONICET} {\normalsize\rm and}
Fidel A. Schaposnik, \thanks
	{Investigador CICBA, Argentina}\\
{\normalsize\it
Departamento de F\'\i sica, Universidad Nacional de La Plata,}\\
{\normalsize\it
C.C.~67, (1900) La Plata, Argentina.}
}
\date{}

%\def\thepage{\protect\raisebox{0ex}{\ } La Plata 94-09}
%\thispagestyle{headings}
%\markright{\thepage}
%\pagenumbering{arabic}

\maketitle
\begin{abstract}
We discuss bosonization in three dimensions of an $SU(N)$ massive
Thirring model in the low-energy regime.  We find that the bosonized
theory is related (but not equal) to $SU(N)$ Yang-Mills-Chern-Simons
gauge theory.  For free massive fermions bosonization leads, at
low energies, to the pure $SU(N)$ (level $k=1$) Chern-Simons theory.
\end{abstract}
\newpage
%===================================================================

\section{Introduction}
\label{sec:intro}

In this paper we investigate the problem of identifying a bosonic
equivalent of a theory of self-interacting fermions with $SU(N)$
symmetry in $2+1$ dimensions.  In a previous publication~\cite{FS},
two of us showed that the low-energy sector of the $U(1)$ massive
Thirring model in $2+1$ dimensions is equivalent to the
Maxwell-Chern-Simons gauge theory.  Here we extend that analysis to
a theory with non-abelian $SU(N)$ symmetry.  We show that, just as
in the abelian case, it is possible to bosonize the low-energy regime
of the theory.  It is also a gauge theory and, as expected, it is
non-abelian and it is closely related to a (level $k=1$) $SU(N)$
Chern-Simons gauge theory.  Naively, one would have expected that
in the non-abelian case the bosonic theory should be the
Yang-Mills-Chern-Simons (YMCS) theory.  Surprisingly, we find that
the bosonized theory is related, but not identical, to YMCS.  However,
in the limit of weak (fermionic-) coupling $g \to 0$, the bosonic
theory becomes equal to the CS gauge theory.  At first sight it may
seem surprising that a simple theory such as free massive fermions
may be equivalent to YMCS.  It may be noted that also in $1+1$
dimensions the bosonized theory of a massive free fermion is also
non-trivial: it is equivalent to a special case of the sine-gordon
theory~\cite{coleman,mandelstam} (in the abelian case) and to a
perturbed Wess-Zumino-Witten theory~\cite{witten84}.

We will follow the same strategy as in reference~\cite{FS} and seek a
bosonic theory which reproduces correctly the {\it low-energy\/}
regime of the massive fermionic theory.  It should be stressed that
this procedure is, in a sense, opposite to what is done in $1+1$
dimensions. There, bosonization is a set of operator identities
valid at length scales {\it short\/} compared with the Compton
wavelength of the fermions.  Here, instead, we only consider the
long distance regime.

This paper is organized as follows. In Section \ref{sec:mapping} we present the
mapping of the low energy sector of the non-abelian fermionic theory into an
equivalent gauge theory. In Section \ref{sec:identities} we derive a set of
identities
for the fermionic currents. In Section \ref{sec:wilson} we discuss the role of
the
Wilson loops of the gauge theory in the equivalent Fermi theory. In
Section\ref{sec:conclusions} we draw a few conclusions on the mapping presnted
here.

\section{The Mapping}
\label{sec:mapping}

We start from the three-dimensional (Euclidean) massive $SU(N)$
Thirring model Lagrangian:
\beq
{\cal L}_{Th} =
 \bar\psi (\id + m) \psi -\frac{g^2}{2}j^{a\mu}j_{\mu}^a
\label{1}
\eeq
where $\psi$ is a two-component Dirac spinor in the fundamental
representation of $SU(N)$, and $j^{a\mu}$ the $SU(N)$ current,
\beq
j^{a\mu} = \bar\psi^i t^a_{ij} \gamma^{\mu} \psi^j.
\label{2}
\eeq
The coupling constant $g^2$ has dimensions of inverse mass.
(Although non-renormalizable by power counting, four fermion
interaction models in $2+1$ dimensions are known to be
renormalizable in the $1/N$ expansion~\cite{Gross}.) We normalize
the $SU(N)$ generators according to $\tr t^a t^b = \delta^{ab}$.

The partition function for the theory is defined as
\beq
\Z_{Th} = \int \D\bar\psi\D\psi
  \exp\{
 -\!\int\dv[\bar\psi (\id + m) \psi -\frac{g^2}{2}j^{a\mu} j_{a\mu}]
  \}  .
\label{3}
\eeq
We eliminate the quartic interaction by introducing a vector field
$a_{\mu}$ taking values in the Lie algebra of $SU(N)$, through the
identity
\beq
\exp\{\frac{g^2}{2} \int\dv\, j^{a\mu}j_{\mu}^a \} =
  \int\D a_{\mu}
  \exp
  \{
    -\tr\int\dv(\frac{1}{2g^2}a^{\mu}a_{\mu} + j^{\mu}a_{\mu})
  \}
\label{4}
\eeq
(up to a multiplicative normalization constant), so that the partition
function becomes
\beq
\Z_{Th} =  \int\D \bar\psi \D\psi \D a_{\mu}
  \exp\{
    -\!\int\dv[ \bar\psi(\id + m + \as)\psi +
	        \frac{1}{2g^2}\tr a^{\mu}a_{\mu}
		    ]
  \}  .
\label{5}
\eeq

The fermionic path-integral can now be done and it yields, as usual,
the determinant of the Dirac operator
\beq
\int \D\bar\psi \D\psi \exp\{
  -\!\int\dv \bar\psi (\id + m + \as) \psi \} = \det (\id + m + \as)
\label{8}
\eeq
Being the Dirac operator unbounded, its determinant requires
regularization. Any sensible regularization approach (for example,
$\zeta$-function or Pauli-Villars) gives a parity violating
contribution~\cite{NS}-\cite{GMSS}. There are also parity conserving
terms which have been computed as an expansion in inverse powers of
the fermion mass:
\beq
\ln \det (\id + m + \as) =
  \pm \frac{1}{16\pi} S_{CS}[a] +
  I_{PC}[a] +
  O(\d^2/m^2)  ,
\label{9f}
\eeq
where the Chern-Simons action $S_{CS}$ is given by
\beq
  S_{CS}[a] = \int\dv
i \epsilon_{\mu\nu\lambda} \tr \int\dv
  (
   f_{\mu \nu} a_{\lambda} -
   \frac{2}{3} a_{\mu}a_{\nu}a_{\lambda}
  )  .
\eeq
Concerning the parity conserving contributions, one has
\beq
I_{PC}[a] =
  - \frac{1}{24\pi  m} \tr\int\dv f^{\mu\nu} f_{\mu\nu}
  + \cdots  ,
\label{8f}
\eeq
where
\beq
f_{\mu\nu} =
  \d_{\mu}a_{\nu} -
  \d_{\nu}a_{\mu} +
  [a_{\mu},a_{\nu}]  .
\label{curv}
\eeq

Using this result we can write $Z_{Th}$ in the form
\beq
Z_{Th} = \int \D a_{\mu}  \exp(-S_{eff}[a])  ,
\label{12f}
\eeq
where $S_{eff}$ is given by
\beq
S_{eff}[a] =
  \frac{1}{2g^2} \tr\int\dv a_{\mu}a^{\mu}  \mp
  \frac{1}{16\pi}S_{CS}[a] +
  \frac{1}{24\pi m} \tr\int\dv f^{\mu\nu} f_{\mu\nu} +
  O(\d^2/m^2)  .
\label{thicon}
\eeq
Up to corrections of order $1/m$, we recognize in $S_{eff}$ the
non-abelian version of the self-dual action $S_{SD}$ introduced
some time ago by Townsend, Pilch and van Nieuwenhuizen~\cite{TPv},
\beq
S_{SD}[a] = \int\dv
  \frac{1}{2g^2} \tr a_{\mu}a^{\mu} \mp
  \frac{1}{16\pi}S_{CS}[a]   .
\label{SD}
\eeq
Then, to leading order in $1/m$ we have established the following
identification:
\beq
  Z_{Th} \approx  Z_{SD} = \int \D a_{\mu} \exp(-S_{SD}[a])
\label{nue}
\eeq

In the abelian case, it has been shown by Deser and Jackiw~\cite{DJ}
that the model with dynamics defined by $S_{SD}$ is equivalent to
the Maxwell-Chern-Simons (MCS) theory.  Using this connection, we
have shown in \cite{FS} the equivalence of the abelian Thirring
model and the MCS theory.  Our proof was based in the use of an
``interpolating Action'' $S_I$ connecting the self dual and
the MCS actions.  It has been already recognized in~\cite{DJ} that
the non-abelian extension of these kind of equivalences is more
involved.  In fact, we shall show here that the non-abelian self-dual
action (and, consequently, the $SU(N)$ Thirring model) is not
equivalent to a Yang-Mills-Chern-Simons theory (the natural extension
of the abelian MCS theory) but to a model where, instead of the
Yang-Mills action, one has a more complicated term, which reduces
to the quadratic $\tr F_{\mu\nu}^2$ only in the $g^2 \to 0$ limit.

To see this, we consider, following \cite{LPvT}, the interpolating
Lagrangian
\beq
L_I[a,A] =
  \frac{1}{8g^2} \tr a_{\mu}^2
  \pm \frac{i}{16\pi} \epsilon_{\mu\nu\lambda}\tr a_\mu
      (F_{\nu\lambda} + A_\nu a_\lambda)
  \pm \frac{1}{16\pi} L_{CS}[A]  ,
\label{inter}
\eeq
with $F_{\mu\nu}$ the non-Abelian curvature
\beq
F_{\mu \nu} =
  \d_\mu A_\nu - \d_\nu A_\mu + [A_\mu,A_\nu]  .
\label{ff}
\eeq
and $ L_{CS}[A]$ is the Chern-Simons Lagrangian.
The partition function associated to $L_I$ is defined as
\beq
Z_I = \int \D A_{\mu}\D a_{\mu} \exp\{-\!\int\dv L_I[a,A]\} .
\label{partit}
\eeq

In order to see the connection between $Z_I$ and the partition
function $Z_{SD}$ for the self-dual system, we shift
in~(\ref{partit}) the $A_\mu$ variables as follows:
\beq
A_\mu = {\bar A}_\mu - a_\mu  .
\label{shift}
\eeq
Under this change, one has
\beq
F_{\mu\nu}[A] =
  F_{\mu\nu}[\bar A] - D[\bar A]_{[\mu}a_{\nu]} + [a_\mu,a_\nu]  ,
\label{f}
\eeq
with the covariant derivative defined as
\beq
D_\mu[\bar A] = \d_\mu + [{\bar A}_\mu,\ ]  .
\label{cov}
\eeq
Concerning the Chern-Simons Lagrangian, one has
\beq
L_{CS}[A] =
  L_{CS}[\bar A] -
  2 \epsilon_{\mu\nu\lambda} a_\mu
    (F_{\nu\lambda}[\bar A] - D_\nu[\bar A]a_\lambda +
     \frac{2}{3} a_\nu a_\lambda)  .
\label{shift1}
\eeq
Puting all this together we easily find
\beq
L_I[a,A] =
  \frac{1}{2g^2} \tr a_\mu a^\mu \pm
  \frac{1}{16\pi} L_{CS}[\bar A] \mp
  i\frac{1}{8\pi} \tr \epsilon_{\mu\nu\lambda}\,
    a_\mu(\d_\nu a_\lambda + \frac{2}{3} a_\nu a_\lambda)  .
\label{final}
\eeq
Transformation~(\ref{shift}) has unit Jacobian and completely
decouples the $\bar A$ field. Denoting by $N$ the result of
integrating over $\bar A$, one then has
\begin{eqnarray}
Z_I
& = &
  N \int \D a_\mu
  \exp\{
    -\frac{1}{2g^2} \tr\int\dv a_\mu a^\mu
  \}
  \times\nonumber\\
& &
  \exp\{
    \pm i\frac{1}{8\pi} \tr\int\dv
	  \epsilon_{\mu\nu\lambda}a_\mu
      (\d_\nu a_\lambda + \frac{2}{3}a_\nu a_\lambda)
  \}  .
\label{1eq}
\end{eqnarray}
We recognize in this expression the action for the  self-dual system
defined in eqs.~(\ref{SD}).  Then, we have proven that
\beq
Z_I = N \int \D a_\mu \exp(-S_{SD})  .
\label{pufff}
\eeq
Comparing eqs.~(\ref{nue}) and~(\ref{pufff}) we can establish the
following relation
\beq
Z_{Th} \approx Z_I  .
\label{ahora}
\eeq

We shall now proceed to perform the path-integrations in $Z_I$ in
the inverse order, that is, first integrating over $a_\mu$.  To
this end, starting again from eq.~(\ref{inter}) we write
\beq
Z_I =
  \int \D A_\mu \exp(\mp \frac{1}{16\pi}S_{CS}[A]) \times \exp(-I[A])  ,
\label{I}
\eeq
with
\begin{eqnarray}
\exp(-I[A])
& = &
  \int \D a_\mu \exp\{-\frac{1}{8g^2} \tr\int\dv a_\mu a^{\mu}\}
  \times\nonumber \\
&  &
  \exp \{ \mp i\frac{1}{16\pi} \tr\int\dv
    \epsilon_{\mu\nu\lambda}
	a_\mu (F_{\nu\lambda} + A_\nu a_\lambda) \}  .
\label{II}
\end{eqnarray}
We can rewrite $I[A]$ in the form
\beq
\exp(-I[A]) =
  \int \D a_\mu
  \exp\{
    - \tr\int\dv
    (
	  \frac{1}{2} a_\mu S^{-1}_{\mu\nu} a_\nu
	  \pm \frac{i}{8\pi} \Fd_\mu a_\mu
	)
  \}  ,
\label{III}
\eeq
with
\beq
 S^{-1 ab}_{\mu\nu} =
   \frac{1}{4g^2} \delta_{\mu\nu}\delta^{ab}
   \mp \frac{i}{8\pi} \epsilon_{\mu\nu\lambda} f^{acb} A^c_\lambda
\label{IV}
\eeq
and
\beq
\Fd_{\mu} =
  \frac{1}{2}\epsilon_{\mu\nu\lambda} F_{\nu\lambda}  .
\label{dual}
\eeq

Then, if we perform the path-integral over $a_\mu$ we find
\beq
I[A] =
  \frac{1}{128\pi^2} \tr\int\dv\Fd_\mu S_{\mu \nu}\Fd_\nu  .
\label{IA}
\eeq
Hence, we see that $Z_I$ can be written in the form
\beq
Z_I = \int \D A_\mu \exp(-S_{FCS}[A]) \equiv Z_{FCS}
\label{bas}
\eeq
where
\beq
S_{FCS}[A] = \pm \frac{1}{16\pi} S_{CS}[A] +
  \frac{1}{128\pi^2} \tr\int\dv\Fd_\mu S_{\mu \nu}\Fd_\nu
\label{algos}
\eeq
Then, using eq.~(\ref{ahora}) we have the relation
\beq
Z_{Th} \approx Z_{FCS}
\label{alfin}
\eeq
which shows the equivalence (to order $1/m$) of the fermionic
Thirring model and a bosonic theory with action $S_{FCS}$.  Let
us notice that only in the limit $g^2 \to 0$ the action $S_{FCS}$
reduces to the Yang-Mills-Chern-Simons action.

\section{Bosonization Identities}
\label{sec:identities}

In order to infer the bosonization identities which derive from the equivalence
found
in the last section,
we add a source for the Thirring current:
\beq
L_{Th}[b_\mu] = L_{Th} + \tr \int\dv j^\mu b_\mu
\label{s1}
\eeq
Then, instead of~(\ref{5}), the partition function now reads
\beq
\Z_{Th}[b] =
  \int \D\bar\psi \D\psi \D a_\mu
  \exp\{
    -\int\dv [\bar\psi (\id + m + \as + \bs)\psi
    +\frac{1}{2g^2} \tr a^\mu a_\mu]
  \}  .
\label{s5}
\eeq
Or, after shifting $a_\mu \to a_\mu - b_\mu$,
\beq
Z_{Th}[b] =
  \exp \{-\frac{1}{2g^2} \tr \int\dv b_\mu b^\mu\} \times
  \int \D a_\mu
  \exp\{-S_{eff}[a] + \frac{1}{g^2} \tr \int\dv b_\mu a^\mu\}
\label{s6}
\eeq
with $S_{eff}[a]$ still given by eq.~(\ref{thicon}).  Then, we can
again establish to order $1/m$ the connection between the Thirring
and self-dual models, in the presence of sources:
\beq
Z_{Th}[b] =
  \exp \{ -\frac{1}{2g^2} \tr \int\dv b_\mu b^\mu \} \times
  \int \D a_\mu
    \exp\{ -S_{SD}[a] + \frac{1}{g^2} \tr \int\dv b_\mu a^\mu\}  .
\label{s7}
\eeq
or
\beq
Z_{Th}[b] =
  \exp \{ -\frac{1}{2g^2} \tr \int\dv b_\mu b^\mu \} \times
  Z_{SD}[b]  ,
\label{s8}
\eeq
where
\beq
Z_{SD}[b] =
  \int\D a_\mu
  \exp\{ -S_{SD}[a] + \frac{1}{g^2} \tr\int\dv b_\mu a^\mu \}
\label{demas}
\eeq
In order to connect this with the vector boson system, let us
again consider the interpolating Lagrangian $L_I$ (eq.~\ref{inter}),
but now in the presence of sources:
\beq
L_I[a,A,b] = L_I[a,A] + \frac{1}{4g^2} \tr a_\mu b^\mu  .
\label{inter1}
\eeq
By shifting as before $A_\mu \to A_\mu - a_\mu$ and integrating out
$A_\mu$, one easily shows that the corresponding partition function
$Z_I [b]$ coincides (up to a normalization factor) with  $Z_{SD}[b]$,
\beq
Z_{SD}[b] = Z_{I}[b]
\label{ag}
\eeq
and then
\beq
Z_{Th}[b] = \exp\{-\frac{1}{2g^2} \tr \int\dv b^\mu b_\mu\}
            \times Z_I[b]  .
\label{zee}
\eeq
If we integrate in the inverse order we have (after shifting
$a_\mu \to 2 a_\mu$),
\beq
Z_I[b] =
  \int\D A_\mu\exp(\mp \frac{1}{16\pi} S_{CS}[A]) \times
  \exp(-I[A,b])
\label{I1}
\eeq
with
\beq
\exp(-I[A,b]) =
  \int \D a_\mu
  \exp \{
    - \tr\int\dv
	[
	  \frac{1}{2}a_\mu S^{-1}_{\mu\nu}a_\nu \pm
	  \frac{i}{8\pi}(\Fd_\mu \mp i \frac{8\pi}{g^2}b_\mu) a_\mu
	]
  \}  .
\label{IIIa}
\eeq
This means that $Z_I[b]$ can be written in the form
\begin{eqnarray}
Z_I[b] & = &
  \int \D A_\mu \exp(\mp \frac{1}{16\pi}S_{CS}[A]) \times
\nonumber \\
& &
  \exp \{
    -\frac{1}{128\pi^2}\tr\int\dv
	(\Fd_\mu \mp i\frac{8\pi}{g^2}b_\mu) S_{\mu\nu}
	(\Fd_\nu \mp i\frac{8\pi}{g^2}b_\nu)
  \}  ,
\label{lio}
\end{eqnarray}
so we have the relation
\begin{eqnarray}
Z_{Th}[b] & = &
  \exp\{-\frac{1}{2g^2} \tr\int\dv b^\mu b_\mu \} \times
  \int\D A_\mu \exp(-S_{FCS}[A]) \times \label{uf} \\
& &
  \exp\{\frac{1}{8g^4} \tr\int\dv b^\mu S_{\mu\nu} b^\nu \}
  \times\exp\{\pm\frac{i}{16\pi g^2}\tr\int\dv b^\mu S_{\mu\nu}\Fd^\nu\}  .
\nonumber
%\label{uf}
\end{eqnarray}
This is, in its most general form, the result we were after.
It provides a complete low-energy bosonization prescription,
valid for any $g^2$, of the matrix elements of the fermionic
current.  Since after differentiating we must set $b_\mu=0$,
we see that, as suggested by eq.~(\ref{alfin}), the bosonized
version of the Thirring model is described by the action
$S_{FCS}$ in eq.~(\ref{algos}).  However, the bosonization
rule for the fermionic current is not simple.  For instance,
from eq.~(\ref{uf}) we get, up to contact terms,
\beq
 j_\mu^a  \to
  \pm \frac{i}{16\pi g^2}  S_{\mu\nu}^{ab} \Fd^\nu_b  ,
\eeq
However, higher derivatives respect to the sources lead to more involved
bosonic equivalents of the vacuum expectation value of products of
fermionic currents.

This more complex structure should not come as a surprise: even in
two-dimensions a simple bosonization procedure applies only to free
fermions \cite{witten84}.  Then, with that in mind, we restrict the
discussion that follows to the $g^2 \to 0$ limit, where with
eq.~(\ref{IV}),  eqs.~(\ref{algos}, \ref{uf}) reduce to
\begin{eqnarray}
Z_{Th}[b] &=&
  \int \D A_\mu
  \exp( \mp\frac{1}{16\pi} S_{CS}[A] ) \times
 \exp\{\pm\frac{i}{4\pi}\tr\int\dv b_\mu \Fd^\mu \}\times \nonumber\\
& & \exp\{\pm\frac{i}{4\pi}\tr\int\dv \epsilon^{\mu\alpha\nu}
b_\mu[A_\alpha,b_\nu]\}  ,
\label{bos1a}
\end{eqnarray}
so, in that limit, the ground state fermionic current maps to
\beq
j_\mu^a \to
  \pm \frac{i}{4\pi} \Fd^a_\mu  ,
\label{bos1b}
\eeq
Concerning the factor in (\ref{bos1a}) which is
quadratic in the sources, its first contribution
will arise when computing current-current correlation functions.
One can see however that the resulting commutator algebra, obtained
via the Bjorken-Johnson-Low method from these correlation
functions, is not modified by the quadratic term. In this
sense, eq.(\ref{bos1b}) gives the bosonization
mapping for non-abelian free fermions in $3$ dimensions as
the natural generalization of the result obtained
in ref.\cite{FS} for the abelian case. Of course
the fact that our results only hold at long distances makes
the analysis of the commutator algebra, which tests short distances,
not completely reliable.

We thus see that
the non-abelian bosonization of free $SU(N)$ massive fermions in
$2+1$ dimensions leads to the (level $k=1$) $SU(N)$ Chern-Simons
theory, with the fermionic current being mapped to the dual of the
gauge field strength.  As stated earlier, this result holds only
for length scales large compared with the Compton wavelength of the
fermion, since our results were obtained for large fermion mass.
It is important to notice that the limit $g^2 \to 0$ to which we
restrict henceforth, corresponds to free fermions but not to an
abelian gauge theory.  On the contrary.  $F^a_{\mu\nu}$ is the
full non-abelian field strength (cf. eq.~(\ref{ff})) and the
Yang-Mills coupling is proportional to $1/g^2 \to \infty$, which
is why we are left with a pure Chern-Simons theory and not with a
mixed Yang-Mills-Chern-Simons action.

Eq.~(\ref{bos1b}) gives a natural non-abelian extension of the
abelian bosonization rule obtained in ref.~\cite{FS}.  In the
abelian case one can interpret the $(2+1)$-dimensional bosonization
formula (which is identical to eq.~(\ref{bos1b}) but with
$F_{\mu\nu}$ the abelian curvature) as the analog of the
$(1+1)$-dimensional result $\bar\psi\gamma^{\mu}\psi \to
({1}/{\sqrt{\pi}}) \epsilon_{\mu\nu}\d^\nu\phi$.
(The factor of $i$ in the expression for the current in
eq.~(\ref{bos1b}) appears because we are working in Euclidean space).
To establish this correspondence also in the non-abelian model,
one should remember that in this last case the two dimensional
bosonization identity reads~\cite{witten84}
\beq
j_+ \to -\frac{i}{4\pi} h^{-1}\d_+ h
\label{w1}
\eeq
\beq
j_- \to -\frac{i}{4\pi} h \d_- h^{-1}
\label{w2}
\eeq
with fermions in the fundamental representation of the group $G$,
and $h$ an element of $G$.

To make  contact with our $(2+1)$-dimensional result, let us first
note that in the $g^2 \to 0$ limit the resulting bosonic action
(the Chern-Simons action) is gauge invariant and so a gauge fixing
is required.  The natural gauge in order to compare the results in
$1+1$ and $2+1$ dimensions is the $A_3 = 0$ gauge. Moreover, one
can write
\beq
A_\pm = A_1 \pm i A_2
\label{cl}
\eeq
in the form
\beq
A_+ =  -i h^{-1} \d_+ h
\label{am}
\eeq
\beq
A_- =  -i f^{-1} \d_- f
\label{ame}
\eeq
With this, one has from~(\ref{bos1b}),
\beq
j_+ \to \pm \frac{i}{4\pi} \d_3( h^{-1} \d_+ h)
\label{jm}
\eeq
\beq
j_- \to \mp \frac{i}{4\pi} \d_3( f^{-1} \d_- f)
\label{jmm}
\eeq
These are the $(2+1)$-dimensional analogs of the two-dimensional
formulas (\ref{w1})-(\ref{w2}). Concerning the additional component
$j_3$, one has
\beq
j_3 =
  \pm \frac{i}{8\pi}
  \left(
    \d_+ ( f^{-1} \d_- f ) -
    \d_- ( h^{-1} \d_+ h ) +
    i ( h^{-1} \d_+h,  f^{-1} \d_- f )
  \right)
\label{j3}
\eeq

\section{Wilson Loops}
\label{sec:wilson}

Perhaps the most interesting aspect of the bosonization identities
of eqs. (\ref{bos1a})-(\ref{bos1b}) is the promotion of the global
$SU(N)$ symmetry of the free fermions to a local gauge symmetry
in the bosonic theory.  To explore the contents of this bosonization
rule we consider the natural objects in the gauge theory, namely
the vacuum expectation value of Wilson loops.  In the Chern-Simons
theory they measure topological invariants determined by the linkings
of the loops and by the topology of the base manifold \cite{witten89}.
For one loop $\Gamma$,
\beq
W[\Gamma] =
  \tr P \exp(i \oint_\Gamma dx^\mu\,A_\mu )
\label{wloop}
\eeq
where $P$ denotes the path ordering of the exponential, and the trace
is taken in the representation carried by the loop.  According to
the bosonization prescription of eq.~(\ref{bos1b}), to relate this
operator to the fermionic theory we must express $W[\Gamma]$ in
terms of the field strength $F_{\mu\nu}$ rather than the potential
$A_\mu$.  In the abelian case this can always be done by means
of Stokes theorem.  As discussed in ref.~\cite{FS}, this leads to
an explicit mapping between abelian Wilson loops and non-local
fermionic operators. Hence, in this way, the latter are related to
the linking of loops and thus probe the generalized statistics of
the external particles that propagate along those loops.  One way
to extend that analysis to the non-abelian case is to use the
non-abelian extension of Stokes theorem developed in~\cite{bralic}.
For an arbitrary loop $\Gamma = \d\Sigma$, the boundary of a surface
$\Sigma$, one has
\beq
W[\d\Sigma] = \tr P_t \exp\{ i\!\int_0^1 dt \int_0^1 ds
  \frac{\d\Sigma^\mu}{\d s}\frac{\d\Sigma^\nu}{\d t}
  W^{-1}[_s\Sigma(t)_0] F_{\mu\nu}(\Sigma(t,s)) W[_s\Sigma(t)_0] \}
\label{stokes}
\eeq
Here $\Sigma$ is looked upon as a sheet, that is, a one parameter
family of paths parametrized by $t$, $0 \leq t \leq 1$.  For each
$t$, $\Sigma(t)$ is a path, itself parametrized by $s$,
$0 \leq s \leq 1$, with fixed end-points:  $\d\Sigma(t,s)/\d t = 0$
at $s = 0,1$.  For a given $t$, $_s\Sigma(t)_0$ denotes the segment
of the path $\Sigma(t)$ connecting the points $\Sigma(t,0)$ and
$\Sigma(t,s)$, and $W[_s\Sigma(t)_0]$ is the corresponding (open)
Wilson line.  Finally, $P_t$ in eq.~(\ref{stokes}) denotes ordering
of the $t$ integral, while the $s$ integral is not ordered (although
there is an $s$-ordering inside each $W[_s\Sigma(t)_0]$.)

In the abelian case the two open Wilson lines $W[_s\Sigma(t)_0]$
in eq.~(\ref{stokes}) cancel each other and one recovers the usual
Stokes theorem, involving only the gauge field strength.  In the
non-abelian case, however, the factors $W[_s\Sigma(t)_0]$ are needed
for gauge invariance, and introduce an explicit dependence of the
Wilson loop operator on the gauge potential $A_\mu$.  Thus, as
opposed to the abelian case, the non-abelian Wilson loop operator
cannot be mapped in a straightforward way to a fermionic operator
through the bosonization rule in eq.~(\ref{bos1b}).

For planar loops this difficulty is only apparent.  Indeed, consider
$W[\d\Sigma]$, with $\Sigma$ contained, say, in the $(1,2)$ plane.
Imposing the $A_3 = 0$ gauge condition, there is a remnant gauge
freedom for the $A_1$ and $A_2$ components in the $(1,2)$ plane,
which is the symmetry of a $2$-dimensional gauge theory in that
plane.  As discussed in~\cite{bralic}, one can use that gauge
symmetry, together with the freedom of parametrization of the
surface $\Sigma$, so the open Wilson line elements in the right hand
side of eq.~(\ref{stokes}) become the identity.  More precisely,
choosing the gauge condition $A_2 =0$ on the $\Sigma$-plane,
eq.~(\ref{stokes}) can be simplified to
\beq
W[\d\Sigma] = \tr P_t \exp\{ i\!\int_0^1 dt \int_0^1 ds
  \frac{\d\Sigma^\mu}{\d s}\frac{\d\Sigma^\nu}{\d t}
  F_{\mu\nu}(\Sigma(t,s)) \}
\label{stokes-plane}
\eeq
provided that $\Sigma$ is parametrized so as to have $\d\Sigma/\d t$
and $\d\Sigma/\d s$ parallel to the $x_1$ and $x_2$ axis,
respectively.  This apparent breaking of rotational invariance,
which includes the presence of $t$-ordering but not of $s$-ordering,
is a consequence of the $A_2 =0$ gauge condition on the
$\Sigma$-plane, and will be removed by the functional integral over
the gauge fields.  Then, writing
\beq
\J[\Sigma] =
  \tr P_t \exp\{\pm 4\pi\!\int_0^1 dt \int_0^1 ds
  \frac{\d\Sigma^\mu}{\d s}\frac{\d\Sigma^\nu}{\d t}
  \epsilon_{\mu\nu\lambda} j_\lambda[\Sigma(t,s)] \}
\label{wsurface}
\eeq
with $j_\mu$ the fermionic current in eq.~(\ref{2}), the
bosonization formula~(\ref{bos1b}) gives
\beq
\langle \J[\Sigma] \rangle_{ferm} =
  \langle W[\d\Sigma] \rangle_{CS}
\label{bos-plane}
\eeq
where in the left hand side the subindex `ferm' stands for free
fermions.  This is the non-abelian generalization of the result
obtained in~\cite{FS}.  It relates a suitably defined non-abelian
flux of the fermionic current through a flat surface, and the
Wilson loop associated to the boundary of that surface, with both
quantities in the same representation of the group.

It should be stressed that the bosonic side of this relation is,
by definition, independent of the surface $\Sigma$ and its
parametrization.  In the fermionic side, however, this is not
obvious.  The relation was derived assuming a flat surface $\Sigma$,
and it is tempting to assume that this may be extended to smooth
deformations away from the plane.  But more relevant is the apparent
breaking of rotational invariance in the fermionic side due to the
remaining $t$-ordering in eq.~(\ref{wsurface}). This should certainly
be expected to be taken care of by the particular parametrization
assumed above for $\Sigma$.  Indeed, one should expect that the very
need of a parametrization and of a matching ordering of the surface
integral of the fermionic current, is just a limitation of our
present analysis.  In addition, as is well known, the expectation
value of the Wilson loop is singular and must be regularized.  A
natural and consistent regularization scheme is provided by the
framing of the loop~\cite{witten89}.  In the case of a
non-intersecting loop on a plane, considered here, that framing can
be chosen also as a plane loop, not intersecting itself nor the
original loop.  Again, it is not clear at this point how these
singularities in the bosonic side will show up in the (free)
fermionic side, and what role will the framing play from the
fermionic point of view.

It is natural to ask whether this analysis can be extended to
several loops and their possible linkings, as done in~\cite{FS}
for the Abelian case.  In the bosonic side one is interested in
the expectation value
$\langle W[\Gamma_1] W[\Gamma_2] \rangle_{CS}$ or, better yet,
the ratio
\beq
%G[\Gamma_1,\Gamma_2] =
  \frac{\langle W[\Gamma_1] W[\Gamma_2] \rangle_{CS}}
       {\langle W[\Gamma_1] \rangle_{CS}
	    \langle W[\Gamma_2] \rangle_{CS}}
\label{two-loops}
\eeq
For non intersecting loops this is a well defined, non singular
object in the Chern-Simons theory, which depends only on the
linking of the two loops $\Gamma_1$ and $\Gamma_2$~\cite{witten89}.
Assuming this to be non-trivial (and non-singular), the two loops
cannot be flat and lying on the same plane, so the previous
construction fails.  But once the  ratio~(\ref{two-loops}) has
been computed in the Chern-Simons theory, we can take the limit
in which the two loops collapse onto a single plane.  This is a
singular limit in which the loops necessarily intersect each other.
Their linking is not well defined any more, and the value
of~(\ref{two-loops}) depends on the initial non-singular loops
used in the computation.  However, at the classical level, before
the functional integral is performed, we can repeat the previous
construction with no difficulties for any arrangement of loops on
the plane~\cite{bralic}.  Thus, formally we can write
\beq
\frac{\langle \J[\Sigma_1\cup\Sigma_2] \rangle_{ferm}}
     {\langle \J[\Sigma_1] \rangle_{ferm}
      \langle \J[\Sigma_2] \rangle_{ferm}} =
\frac{\langle W[\d\Sigma_1] W[\d\Sigma_2] \rangle_{CS}}
     {\langle W[\d\Sigma_1] \rangle_{CS}
      \langle W[\d\Sigma_2] \rangle_{CS}}
\label{two-surf}
\eeq
where both surfaces $\Sigma_1$ and $\Sigma_2$ are contained in the
same plane.  As we just stated, the bosonic side of this relation
will be ill defined in general.  But it can be given a well defined
meaning by lifting the loops $\d\Sigma_i$ from the plane to non
intersecting three-dimensional loops $\Gamma_i$.  This can be done
in different ways, specifying different linkings of the loops
$\Gamma_i$ compatible with the intersections of their projections
$\d\Sigma_i$ onto the plane.  Correspondingly, in the fermionic side,
the surface $\Sigma_1\cup\Sigma_2$ must be complemented with a
prescription stating the way in which the two surfaces $\Sigma_i$
overlap.  The different possible liftings of the loops specify
different overlaps of the surfaces, as ilustrated in Fig.~(1).
In this way, relation~(\ref{two-surf}) (and its generalizations)
can be viewed as a defining relation, through bosonization, of
the vaccuum expectation value of the flux of the fermionic current
through surfaces with foldings.

\section{Conclusions}
\label{sec:conclusions}

We close with a few remarks on the nature of the mapping discussed
here.  In this paper we showed that the low energy sector of the massive
$SU(N)$ Thirring model in $2+1$
dimensions is equivalent to the long distance regime of a non-abelian gauge
theory,
closely related to the Yang-Mills-Chern-Simons gauge theory. In the weak
coupling
limit, the two theories become identical.
It is worthwile to stress that this mapping only holds
at long distances.  In that regime, the gauge theory is a topological
field theory and so is the fermion theory.  Secondly, just as in the
abelian case, we discover the existence of operators of the Fermi
theory which ought to exhibit fractional statistics.  However, unlike
the abelain theory, these objects are substantially more complex.
Finally, the bosonic theory is, essentially, a level $k=1$ $SU(N)$
Chern-Simons theory.  It would be interesting to find a fermionic
analog of a Chern-Simnos theory with level higher than one.

\vspace{1cm}
\underline{Acknowledgements}  This work was
supported in part by FONDECYT (NB), under Grant No. 751/92,
by
the National Science Foundation under
Grant NSF DMR-91-22385 at the University of Illinois at Urbana
Champaign (EF), by CONICET under Grant  PID 3049/92
(VM, FAS), by the NSF-CONICET
International Cooperation Program through the grant NSF-INT-8902032
and by Fundaciones Andes and Antorchas, under Grant No. 12345/9.
FAS thanks the Universidad Cat\'olica de Chile for its kind hospitality.

%%%%%%%%%%%%%%%%%%%%%%%%%%%%%%%%%%%%%%%%%%%%%%%%%%%%%%%%%%%%%%%%%%%%%%%%%%

%%%%%%%%%%%%%%%%%%%%%%%%%%%%%%%%%%%%%%%%%%%%%%%%%%%%%%%%%%%%%%%%%%%%%%%%%%
\newpage
\begin{figure*}
\caption{Different overlaps of the surfaces $\Sigma_1$ and $\Sigma_2$
on a plane, determined by the possible liftings of the loops
$\d\Sigma_1$ and $\d\Sigma_2$ away from the plane.}
\end{figure*}
%%%%%%%%%%%%%%%%%%%%%%%%%%%%%%%%%%%%%%%%%%%%%%%%%%%%%%%%%%%%%%%%%%%%%%%%%%
\end{document}